\documentclass[useAMS,usenatbib]{mn2e}

\usepackage{graphicx}

\usepackage{ifpdf}

\usepackage{times,color,amssymb}

\usepackage{booktabs}

\usepackage{aas_macros}

\ifpdf
	\usepackage{epstopdf}
\fi

\usepackage{amsmath}

\title[Transition disc evolution]{The long-term evolution of photoevaporating transition discs with giant planets}
\author[G. P. Rosotti et al.]{Giovanni P. Rosotti\thanks{E-mail:
rosotti@ast.cam.ac.uk}$^{1,2,3}$, Barbara Ercolano$^{3,4}$,  James E. Owen$^{5,6}$\\
$^{1}$Institute of Astronomy, Madingley Rd, Cambridge, CB3 0HA, UK\\
$^{2}$Max-Planck-Institut f\"ur extraterrestrische Physik, Giessenbachstra\ss{}e, D-85748 Garching, Germany\\
$^{3}$ Universitats-Sternwarte M\"unchen, Scheinerstra\ss{}e 1, D-81679 M\"unchen, Germany\\
$^{4}$ Excellence Cluster Universe, Boltzmannstr. 2, D-85748 Garching, Germany\\
$^{5}$ Institute of Advanced Study, Einstein Drive, Princeton NJ, 08540, USA\\
$^{6}$ Hubble Fellow}

\begin{document}

\date{Accepted 2015 September 08. Received 2015 September 08; in original form 2015 April 15}

\pagerange{\pageref{firstpage}--\pageref{lastpage}} \pubyear{2013}

\maketitle

\label{firstpage}

\begin{abstract}

Photo-evaporation and planet formation have both been proposed as mechanisms responsible for the creation of a transition disc. We have studied their combined effect through a suite of 2d simulations of protoplanetary discs undergoing X-ray photoevaporation with an embedded giant planet. In a previous work we explored how the formation of a giant planet triggers the dispersal of the inner disc by photo-evaporation at earlier times than what would have happened otherwise. This is particularly relevant for the observed transition discs with large holes and high mass accretion rates that cannot be explained by photo-evaporation alone. In this work we significantly expand the parameter space investigated by previous simulations. In addition, the updated model includes thermal sweeping, needed for studying the complete dispersal of the disc. After the removal of the inner disc the disc is a non accreting transition disc, an object that is rarely seen in observations. We assess the relative length of this phase, to understand if it is long lived enough to be found observationally. Depending on the parameters, especially on the X-ray luminosity of the star, we find that the fraction of time spent as a non-accretor greatly varies. We build a population synthesis model to compare with observations and find that in general thermal sweeping is not effective enough to destroy the outer disc, leaving many transition discs in a relatively long lived phase with a gas free hole, at odds with observations. We discuss the implications for transition disc evolution. In particular, we highlight the current lack of explanation for the missing non-accreting transition discs with large holes, which is a serious issue in the planet hypothesis.

\end{abstract}


\section{Introduction}

Disc dispersal is a crucial problem in understanding planetary formation. In particular, the time for the survival of gas in protoplanetary discs sets important constraints on the time available for the formation of gas rich planets. This is particularly relevant for  the core accretion scenario, where the timescale of giant planet formation \citep{Pollack96} is predicted to be of the same order as the lifetime of discs.

Discs have a typical lifetime of a few Myrs \citep{Haisch2001,Mamajek2009,Fedele2010,Ribas2014}. For most of their time, the redistribution of angular momentum dominates their evolution and a simple analytical relation \citep[e.g.][]{LyndenBellPringle} can be used to describe the observed decline of the mass accretion rate with the age of the system (e.g. \citealt{Hartmann98}) and their surface density profiles \citep{Andrews09}. The exact nature of the source of viscosity is still matter of debate, although the magnetorotational instability is the most plausible candidate \citep{PPVIAngMom}. Observations of star forming regions also reveal a particular class of discs, called ``transition discs'' \citep{Strom1989,Skrutskie1990,Espaillat2014}. These discs show a dip in their spectral energy distributions (SED) at near-infrared (NIR) wavelengths, which is usually interpreted as a deficit of warm dust in their innermost regions (i.e., a few AUs) \citep{Calvet2005,Currie2009,Muzerolle2010,Lada2006,Cieza2010,Espaillat2010}. This is not expected in the viscous evolution framework, and it shows that, while viscosity is certainly a main ingredient of disc evolution, another process must also be at play. This is further reinforced by the fact that the viscous time-scale in the outer regions of protoplanetary discs ($\gtrsim$100 AU) is $\gtrsim$ Myr, which implies that some other process must drive disc dispersal at these radii. Observations currently show that approximately 10 percent of all observed discs \citep{KenyonHartmann95,Koepferl2013} are  ``transition'' discs (TD). If all discs go through the transition disc phase, this implies that this phase must be short-lived (of order of $10^5$ years), implying that disc evolution observes a ``two-time scale behaviour''. Transition discs have also been imaged by sub-mm interferometers \citep{Pietu2006,Brown2009,Andrews2011,Isella2012}, confirming the presence of huge (tens of AU) cavities in the mm-dust. 


Photoevaporation \citep{Hollenbach1994,Font2004,Alexander2006Hydro} is an ideal candidate to explain, together with viscous evolution, the observed IR colour evolution of discs. \citet{Clarke2001} showed that when the mass accretion rate becomes comparable with the mass loss by photoevaporation, a gap opens at around a few AUs, rapidly shutting down accretion (the so-called UV switch) and producing a transition disc. The outer disc left is quickly eroded from the inside out by photoevaporation, which agrees with observational findings \citep{Ercolano2011,Koepferl2013}. While this explains \textit{some} transition discs, it does not explain \textit{all} of them. In particular, these models predict that discs with large holes should not be accreting, which is at odds with observations. Recent observational work has remeasured accurately the accretion rates of a specific sample of transition discs, selected to be highly accreting and with large holes \citep{Manara2014}, confirming the values of the accretion rates and finding that, from the point of view of accretion, these discs show the same properties of normal discs.

In recent years, other heating mechanisms have been proposed for driving a photo-evaporative wind, namely the far ultra-violet (FUV) and soft X-rays \citep{Ercolano2008,GortiHollenbach2009a,Owen2010}. These models predict significantly higher mass-loss rates, which can create transition discs with higher accretion rates. It is still unclear however which heating mechanism is the dominant one. Direct evidence of the presence of a low-velocity, ionised disc wind has been given through observations of blue-shifted forbidden emission lines such as [NeII] \citep{PascucciSterzik2009,Sacco2012}. The shape of these lines agrees very well with the predictions of photo-evaporation, although it is unfortunately unable to discriminate between the existing models. For what concerns X-ray heating, it has also been recognised that, after hole opening, when the disc is directly exposed to the X-ray radiation from the star, a ``thermal sweeping'' instability can develop, that rapidly destroys the leftovers of the disc \citep{OwenTheory,OwenTS}. This contributes to explaining why non-accreting transition discs with large holes are not observed, as they are destroyed before reaching this state. The revised X-ray-driven rates provide a way of explaining the observed correlation between mass accretion rates and star masses \citep{Ercolano2014}, confirming that photoevaporation is a fundamental ingredient of disc evolution. Roughly 50\% of transition discs can be directly explained by photoevaporation alone \citep{Owen2011,OwenTS} in the X-ray framework, but despite all efforts the subset of strongly accreting TD with large inner holes mentioned above remains unexplained.

This has led to the hypothesis that transition discs might actually consist of two different families \citep{Owen2012}, created by different physical processes. Photoevaporation can explain the class of low accretors, that also happen to have a lower mm flux, while another process would be needed to explain the other ones. Note that in this picture discs belonging to this second class need not necessarily represent the final phases of proto-planetary disc evolution and they do not need to be necessarily short-lived; it is perfectly plausible that only some of the discs are subject to this phase. Specifically, \citealt{Owen2012} speculated that transition discs with large holes and high accretion rates were rare and long lived. The theoretical difficulty in explaining them resides in having to simultaneously explain objects with large holes, almost perfectly clear in the mm resolved images, which sustain a very high mass accretion rate, meaning that they contain large quantities of gas near the central star (although see \citealt{Rosenfeld2014} for an alternative explanation).

Disc gaps carved by giant planets may represent a viable alternative explanation for the class of transition discs with large holes and high accretion rates. If this were true, such discs would represent indirect evidence for the presence of young planets in discs. The model envisaged by \citet{PaardekooperMellema2006} and \citet{Rice2006}, is that a planet, if massive enough to dynamically open a gap in the disc, produces a pressure maximum outside its orbital radius. Dust grains with Stokes numbers $\sim 1$, which for the parameters of such discs correspond to particle sizes $\sim 1 mm$, are trapped at this radius by gas drag on a short ($\ll$ viscous) time-scale. Recent observations support this idea; for example \citet{vanderMarel2013} recently imaged a dust trap in Iras 48 with ALMA. 
While this scenario certainly constitutes a promising avenue, it is still faced with the problem of small grains in the disc. Small grains ($\lesssim 1~\mu$m) are strongly coupled to the gas and do not remain trapped at the pressure maximum, thus they flow through the gap and replenish the inner disc. Even if the mass contained in the small grains is small, they provide most of the opacity in the mid-infrared. For example \citet{Zhu2012} needed to invoke grain growth in the inner disc to remove these small grains which would otherwise produce too much (unobserved) emission in the MIR. Indeed, \citet{ovelar2013} presented models including also the effect of grain coagulation, that in some cases can successfully clear the inner disc of the small dust. Recently, another scenario was proposed by \citet{OwenRadPress}, who investigated the effect of radiation pressure from the accretion luminosity on the planet on the dust grains. In this case, provided there is gas accretion onto the planet, the radiation pressure acts to keep the small dust outside of the gap. The process depends sensitively on the mass of the planet and accretion rate; if this is too small, this effect is not strong enough, and the small dust will flow in the gap. This is consistent with several existing observations, which show a hole at mm wavelength, but have the SED of a primordial disc \citep{Andrews2011}.


The goal of this study is to investigate what evolutionary path may lead to the creation  of transition discs with large holes and high accretion rates. While considerable theoretical effort has been invested in understanding what process may be responsible for the creation of a transition disc, relatively little has been done to understand their time evolution (but see for example \citealt{alexander2009}, although the holes there are much smaller, and \citealt{clarke2013}). In this work we will just assume that a planet is a sufficient condition to produce a transition disc, and we explore what happens when such a system is allowed to evolve. As viscous evolution and photo-evaporation are thought to be the main drivers of disc evolution, these are the ingredients that we consider to follow the evolution in time of the disc. Specifically, we want to address the issue that very few (if any) transition discs with low mass accretion rates and large holes are observed. Naively, one would expect to see the mass accretion rate of these discs going down with time as they age. If this decay happens in a power-law fashion, as it is the case for viscously evolving discs non in transition, then we should see more and more discs at low rates, contrary to the observations.

Thermal sweeping, which is included in our modelling, may explain the deficiency in the observations of weakly transition discs. In this work we shall investigate whether this mechanism is effective enough in destroying the leftovers of transition discs. In a previous paper \citep{Rosotti2013} we studied the evolution of the inner disc, finding that the presence of a planet can trigger its dispersal at times significant earlier than what would have happened otherwise. After the dispersal of the inner disc, the outer disc is directly exposed to the X-ray radiation, and can be prone to the thermal sweeping instability. As this phase is short lived, it is then unlikely to observe it, and we would expect to observe only transition discs that are accreting at higher rates. One goal of this study is to quantify how effective this process may be.

This paper is structured as follows. In section \ref{sec_num_model} we explain our numerical procedure. In section \ref{sec_results} we presents our results, and finally we present our conclusions in section \ref{sec_concl}.

\section{Numerical model}

\label{sec_num_model}

We refer the reader to \citet{Rosotti2013}, where the details of our model are reported. Here we summarise the most important points, and describe the addition of thermal sweeping.

Our model consists of a proto-planetary disc undergoing viscous evolution and X-ray photoevaporation. We follow the evolution in time of the disc using the publicly available hydrodynamical code \textsc{fargo} \citep{Masset2000}, which we modified to include the effect of mass removal by photoevaporation. At a time $t_\mathrm{creation}$ we assume that a planet forms in the disc with an initial mass of $M_p=0.7 M_{jup}$. Similar to many other works found in the literature that studied the long term disc evolution or planet disc interaction \citep[e.g.,][]{nelson2000,armitage2002,alexander2009,Zhu2011}, we do not model the formation of the planet itself. The simulated radial extent of the disc is $[0.1 r_p, 10 r_p]$, where $r_p$ is the planet orbital radius. We employ a resolution of $n_\phi = 256$ cells in the azimuthal direction and $n_r=188$ logarithmically spaced cells in the radial direction, which gives approximately square cells. Our choice of resolution privileges the possibility of running for times comparable to the viscous one. Computational time is saved by modelling the evolution of the disc until the moment of planet insertion in 1d, since the disc is azimuthally symmetric up to that point. Our viscous code uses a flux-conserving donor-cell scheme, implicit in time. Details about the implementation can be found in \citet{2010A&A...513A..79B}. The computational grid of the 1d code covers $[0.0025 \ \mathrm{AU}, 2500 \ \mathrm{AU}]$, and it is comprised of 1000 grid points. The mesh is uniform in a scaled variable $X \propto R^{1/2}$. 

We assume that the planet accretes material from the surrounding circumplanetary disc, as prescribed by \citet{Kley99}. In {\sc fargo} this controlled by a free parameter, $f_\mathrm{acc}$, that expresses the timescale over which the planet accretes material from the surrounding in terms of the dynamical time scale, i.e. $t_\mathrm{acc} =  1/f_\mathrm{acc} \Omega^{-1}$ ). We set $f_\mathrm{acc}=1$ in all of our models, which means that we are working in the ``disc-limited'' regime where the planet is able to accrete all the material the disc gives to it. Such a regime is expected by the core accretion model \citep{Pollack96} after reaching the critical mass, i.e. effectively after becoming a giant planet.

We assume a fixed profile for viscosity and temperature. The sound speed is a fixed function of radius, and is chosen to give a mildly-flaring disc (i.e., $H/R \propto R^{\mathbf{1}/4}$); the normalization is chosen so that at $1 \ \mathrm{AU}$ the aspect ratio $H/R=0.0333$. We evaluate the viscosity using the well-known relation $\nu=\alpha c_s H$ \citep{shakurasunyaev73}, where we set $\alpha=1.5 \times 10^{-3}$. All the discs have the same initial conditions, that are taken from \citet{Owen2011}, who showed that the scatter in the X-ray luminosity alone can account for the scatter in the properties of observed discs. The surface density is assumed to initially have a self-similar profile \citep{LyndenBellPringle}:
\begin{equation}
\Sigma(R,0)=\frac{M_\mathrm{d}(0)}{2 \pi R R_1} \exp (-R/R_1),
\label{eq:sigma}
\end{equation}
where $M_\mathrm{d}(0)$ is the initial mass of the disc and $R_1$ a scale radius describing the exponential taper of the disc's outer region. We set a value of $R_1=18 \ \mathrm{AU}$ and an initial disc mass of 10\% the mass of the star, which is fixed to be $0.7 M_\ast$.

The details of the photoevaporation profile assumed can be found in the appendix of \citet{OwenTheory}. In our simulations we do not allow the planet to migrate, as \citet{Rosotti2013} found migration to have little impact. While we have run a number of extra simulations in the new parameter space to confirm this assumption, due to excessive computational costs our tests can unfortunately not be considered comprehensive. 


\subsection{Thermal sweeping}

By means of 2d hydrodynamic simulations, \citet{OwenTheory} showed that photoevaporating discs are subjected to thermal sweeping, an instability that can rapidly destroy the disc. In particular, thermal sweeping is triggered when the width of the X-ray heated layer becomes comparable with the vertical scale height of the layer. If this condition is satisfied, a slight readjustment in the vertical direction causes the X-rays to penetrate slightly more, heating a wider layer. This causes a runaway effect that was found to effectively destroy the disc on a time-scale $\lesssim 10^{4}$~years. \citet{OwenTS} presented a relation for the critical surface density threshold at the inner edge of the disc for the instability to set in based on the above argument. They found that thermal sweeping sets in when the surface density is lower than:

\begin{eqnarray}
\Sigma_{TS}\!\!\!
& =&  0.14 \,{\rm g}\, {\rm cm}^{-2}\left(\frac{L_X}{10^{30}{\rm \, erg\, s}^{-1}}\right)\left(\frac{T_{1\rm{AU}}}{100{\rm \, K}}\right)^{-1/2}\nonumber\\
&\times&\!\!\!\left(\frac{M_*}{0.7\,{\rm M}_\odot}\right)^{-1/2}\left(\frac{R_{\rm hole}}{10{\rm \, AU}}\right)^{-1/4}\nonumber\\
&\times &\!\!\! \exp\left[\left(\frac{R_{\rm hole}}{10{\rm\, AU}}\right)^{1/2}\left(\frac{T_{1\rm{AU}}}{100{\rm \, K}}\right)^{-1}\right]\label{eqn:ths}
\end{eqnarray}
where $L_X$ is the X-ray luminosity of the star, $T_{1\rm{AU}}$ is the temperature at $1 AU$ and $R_{\rm hole}$ is the size of the hole. We use equation \eqref{eqn:ths} to determine the stability of the outer disc of our simulations. At each snapshot, we monitor if the inner disc is still present or if it has already dispersed. We experimented with using a criterion based on the total column of material in the midplane, ensuring that it is less than $10^{22} cm^{-2}$, which is the penetration depth of $\sim 1 \ \mathrm{keV}$ X-rays, and checking that the mass in the inner disc is a factor of a few (we typically use 5) above the value it had if the disc had the floor surface density everywhere. Apart for some specific cases where manual intervention was required, we find the two criteria to yield little difference in practice. When the inner disc is no longer present, we inspect the maximum of the azimuthally averaged surface density outside the planet orbital radius. If this is smaller than the threshold at that radius, then we assume that the disc is going to be rapidly destroyed by thermal sweeping and we stop the simulation.

\subsection{Parameters varied}

We vary the following parameters:  disc mass (which corresponds to the age of the disc at the moment of planet creation), X-ray luminosity and planet location. Unfortunately we cannot perform a true population synthesis model, as not all the distributions of the parameters are known. In particular, neither the formation time of planets is known, neither the distribution of planet orbital radii, as very little is known about planets at distances of tens of AU. Only recently surveys \citep{Chauvin2014,Brandt2014} are starting to put constraints on the fraction and distribution of planets in wide orbits. However, it is difficult to relate the properties of these planets around main sequence stars to the putative ones present in transition discs, as significant migration might have happened \citep{clarke2013}. Rather, the purpose of this investigation is to get an handle on the possible outcomes in different regions of the parameter space. It is also not clear whether planets can form at all at such large radii. Core accretion models \citep[e.g.,][]{IdaLin2004} predict that planets form closer to the star ($\sim$ 5 AU) where the relevant timescales are short enough that significant growth happens over the disc lifetime. Gravitational instability could be possibly another candidate for the formation of those planets \citep[e.g.,][]{Rafikov2005,Stamatellos2008,Clarke2009}, although no statistical comparison with the theories has been conducted so far to the best of our knowledge. Nevertheless, there is a growing consensus about the presence of planets in transition discs as detailed in the introduction. If observations will prove conclusively that planets really exists at large radii, a revision of the theories of planet formation will be needed to account for this observational evidence. For the time being, we assume the \textit{working hypothesis} that a planet at large radii is responsible for the formation of the cavity in the dust.

We explore values of $4 \times 10^{29}, 7.5 \times 10^{29}, 10^{30}, 2.5 \times 10^{30}, 4 \times 10^{30} \ \mathrm{erg} \mathrm{s}^{-1}$ for the X-ray luminosity. Such values are chosen in order to sample appropriately the Taurus X-ray luminosity function \citep{guedel2007}: the minimum and maximum value chosen are roughly at 16 and 84 percentile, i.e. roughly $1 \sigma$ (if the distribution were gaussian) from the median, which is $10^{30} \ \mathrm{erg} \mathrm{s}^{-1}$. Therefore these values bracket most of the possible outcomes, depending on the level of X-ray luminosity. We experiment with planet locations of $20$ and $40 \ \mathrm{AU}$. For  the disc mass at planet creation, we used values of $15, 30, 45, 60 \ M_\mathrm{jup}$.
\begin{table}
\caption{Parameters varied}
\begin{tabular}{cccc}
Parameter & Values \\ \hline
$M_\mathrm{disc}$ & [15, 30, 45, 60 ] $M_\mathrm{jup}$\\
$R_p$ & [20, 40] $\mathrm{AU}$\\
$L_X$ &  [$4 \times 10^{29}, 7.5 \times 10^{29}, 10^{30}, 2.5 \times 10^{30}, 4 \times 10^{30}] \ \mathrm{erg} \mathrm{s}^{-1}$ \\

\end{tabular}
\end{table}

\section{Results}

\label{sec_results}

\subsection{Classification scheme for comparison with observations}

\label{sec:scheme}

\begin{figure}
\includegraphics[width=\columnwidth]{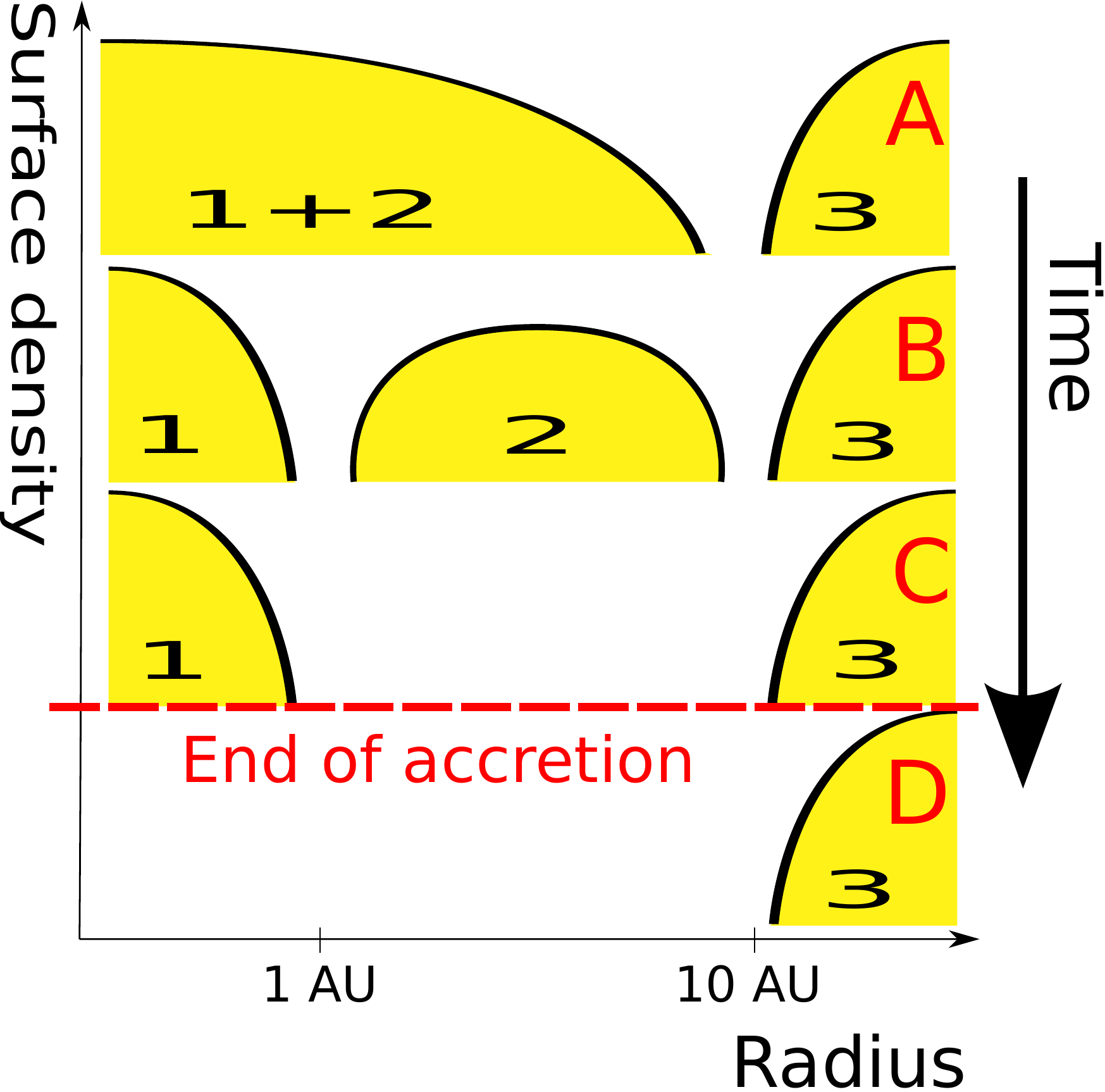}
\caption{Classification scheme used to separate the lifetime of a transition disc in different phases. Phase A and phase B correspond to accreting transition discs with gas in the cavity, phase C is an accreting transition disc without gas in the cavity, and phase D is a non-accreting transition disc without gas in the cavity.}
\label{fig:scheme}
\end{figure}

As already mentioned, we assume that the presence of a planet is a sufficient condition to produce a transition disc and we do not include any modelling of the dust component in our simulations. Therefore, we cannot directly relate the outcomes of our simulations to continuum images or SEDs. Rather, the observational proxies that we would like to compare with are the following:
\begin{itemize}
\item the presence of significant gas in the cavity of transition discs;
\item the presence of a detectable mass accretion rate.
\end{itemize}
We stress that, although we consider the presence of a planet as a \textit{sufficient} condition, we do not consider it as \textit{necessary} for the creation of a transition disc. As remarked in the introduction, roughly half of the known transition discs can be explained by photo-evaporation alone, without the need to invoke the presence of a planet. However, here we are concerned with the discs that \textit{cannot} be explained by photo-evaporation (those with huge cavities and high accretion rates), and that is why we explore the consequences of the presence of a planet.

While some unsuccessful attempt to detect  gas in the cavity was done with the previous generation of interferometers \citep{Dutrey2008,Lyo2011}, ALMA has now the sensitivity and the spatial resolution necessary to accomplish this goal   \citep{RosenfeldTWHya,Bruderer2014,Casassus2013,Zhang2014}. As measurements of this kind are becoming routinely available, it is of primary importance for models of transition discs to provide predictions for the gas content. This quantity is much ``cleaner'' from the hydrodynamical side, in contrast to the dust that can be observed in the continuum. The dust has a much more complicated dynamics, as it interacts differently with the gas depending on its size. In addition, dust grains have a distributions of sizes, and processes of coagulation and fragmentation happening in the disc introduce other uncertainties and make it difficult to predict the dust distribution. However, observations of the gas are also not straightforward, as $H_2$, the most abundant molecule, is essentially invisible. Nevertheless, observations of other molecules (particularly  CO ) and modelling of their chemistry has now advanced to a point where it is possible to compute the expected emission in the gas lines from a given gas surface density. 

It is straightforward to know in our models if CO emission would be observed from a cavity of a transition disc, as the region sampled by ALMA is in our computational grid. \citet{Bruderer2013} showed that ALMA has the sensitivity to detect through observations of the CO 3-2 transition in Band 7 up to a few Earth masses of gas in the cavity. Measuring how much mass is present involves more complicated modelling using different isotopologues of the molecule. For the sake of simplicity, we restrict ourselves here to explore only if there is going to be a detection, rather than focusing on the strength of the expected emission.
Note that  we are more concerned with the detection of the rotational transitions of the CO molecule rather than the observations of the fundamental rovibrational transition at $4.7 \mu m$. The reason is that this transition does not trace the bulk of the gas mass inside the cavity, but rather the warm gas (roughly between 100 and 1000 $K$) up to distances of several AUs.

Unfortunately, it is not straightforward to derive mass accretion rates from our simulations, as the disc inner boundary is not resolved in our simulations. We assume that for the first part of the evolution the inner disc just readjusts its structure in order to supply the star with the same mass accretion rate that we have at the inner boundary of our grid. This assumption is justified by the fact that the viscous time in the inner region that we do not simulate is smaller than the one in the region that we simulate. When photo-evaporation mass-loss rates are coupled with viscous evolution \citep{Clarke2001,Ruden2004,Owen2010}, it is found that when the mass accretion rate goes below the photoevaporation rate, a gap opens in the disc, at a radius around $1 AU$ (for a solar type star). We assume that the clearing of this inner region (that is, disc 1 in Figure \ref{fig:scheme}) is independent of what happens in the outer region of the disc. We thus use 1d models of the whole disc to complement our 2d simulations. We monitor the mass accretion rate at which a gap opens in the disc in our 1d models, and we assume that the same happens in the disc we do not resolve when the mass accretion rate at the inner boundary of the grid reaches this threshold. Subsequently, we take the evolution in time of the mass accretion rate of the 1d model as the mass accretion rate onto the star. Note that this is not the same as what was done by \citet{Rosotti2013}, who matched the 1d and 2d models monitoring the radius of the inner hole in the 2d grid. However, this assumes that the disc resolved in the 2d simulations always clears from inside out. We decided to switch to this new criterion because in some of our simulations, which cover a larger region of the parameter space than what was done by \citet{Rosotti2013}, the central disc clears from the outside in, and therefore the criterion used by \citet{Rosotti2013} would trigger the opening of a photoevaporation gap at slightly later times.

The qualitative behaviour that we have described can be summarised in 4 phases, which are shown in Figure \ref{fig:scheme}. Initially, only the dynamical gap created by the planet is present (phase A), and the disc is otherwise continuous. As the disc ages and mass accretes onto the star, the mass accretion rate will decrease up to the point where photoevaporation is able to open a gap in the innermost region of the disc. We note that, as \citet{Rosotti2013} reported, this will happen at earlier times for discs with a planet, as this partially cuts out the inner from the outer disc, where most of the mass is. Thus, we assume that the disc will have two gaps (phase B), dividing the disc in three regions. Disc 2 is cleared on a relatively short phase (typically less than $10^5$ years) by photoevaporation, leaving only disc 1 and 3 (phase C). This disc will not show emission from the cavity in the CO rotational lines, which trace the bulk of the cold mass, but will still have a measurable accretion onto the star due to the presence of disc A. Finally, disc 1 can not last longer than $~10^5$ years (its viscous time scale), and in the last phase (phase D) only the outer disc will be present. Observationally, this is a non-accreting transition disc without gas in the cavity. Non accreting transition discs are rarely seen in observations, and the main goal of this paper is exactly to quantify how long lived  this phase may be before the set in of thermal sweeping.

As in every classification scheme, we do not expect it to hold necessarily for all discs; rather, we are interested that it catches the majority of the behaviour. We inspected visually the results of our simulations and confirm that this is the case. Differences are found in the way disc 2 clears, which happens sometimes from inside in and sometimes from inside out. Since, as said before, the clearing of this disc is quite fast, this different behaviour is unlikely to have an observational impact.

In 2 simulations (the parameters are $L_X=10^{30} \ \mathrm{erg} \mathrm{s}^{-1}$, $R_p=20 \ \mathrm{AU}$, $M_\mathrm{disc} = 45 M_\mathrm{jup}$ and $L_X = 4 \times 10^{29} \ \mathrm{erg} \mathrm{s}^{-1}$, $R_p=20 \ \mathrm{AU}$, $M_\mathrm{disc}=60 M_\mathrm{jup}$) out of the 40 we have run, disc 2 is eroded briefly from inside out, but then the erosion by photoevaporation reaches an equilibrium with the material filtering through the gap. In one case, disc 2 is very long lived (almost $1 Myr$), creating a transition disc without accretion (as disc 1 cannot be so long lived, being cut out from the mass resupply) but with cold gas in the cavity (although the mass is very small: just a few Earth masses, at the limit of detectability with ALMA). We remove this disc from what follows, as it might falsify the results.


\subsection{Lifetimes}

The main result that we extract from each simulation consists in the length of each phase that was introduced in section \ref{sec:scheme}. 

\begin{figure}
\includegraphics[width=\columnwidth]{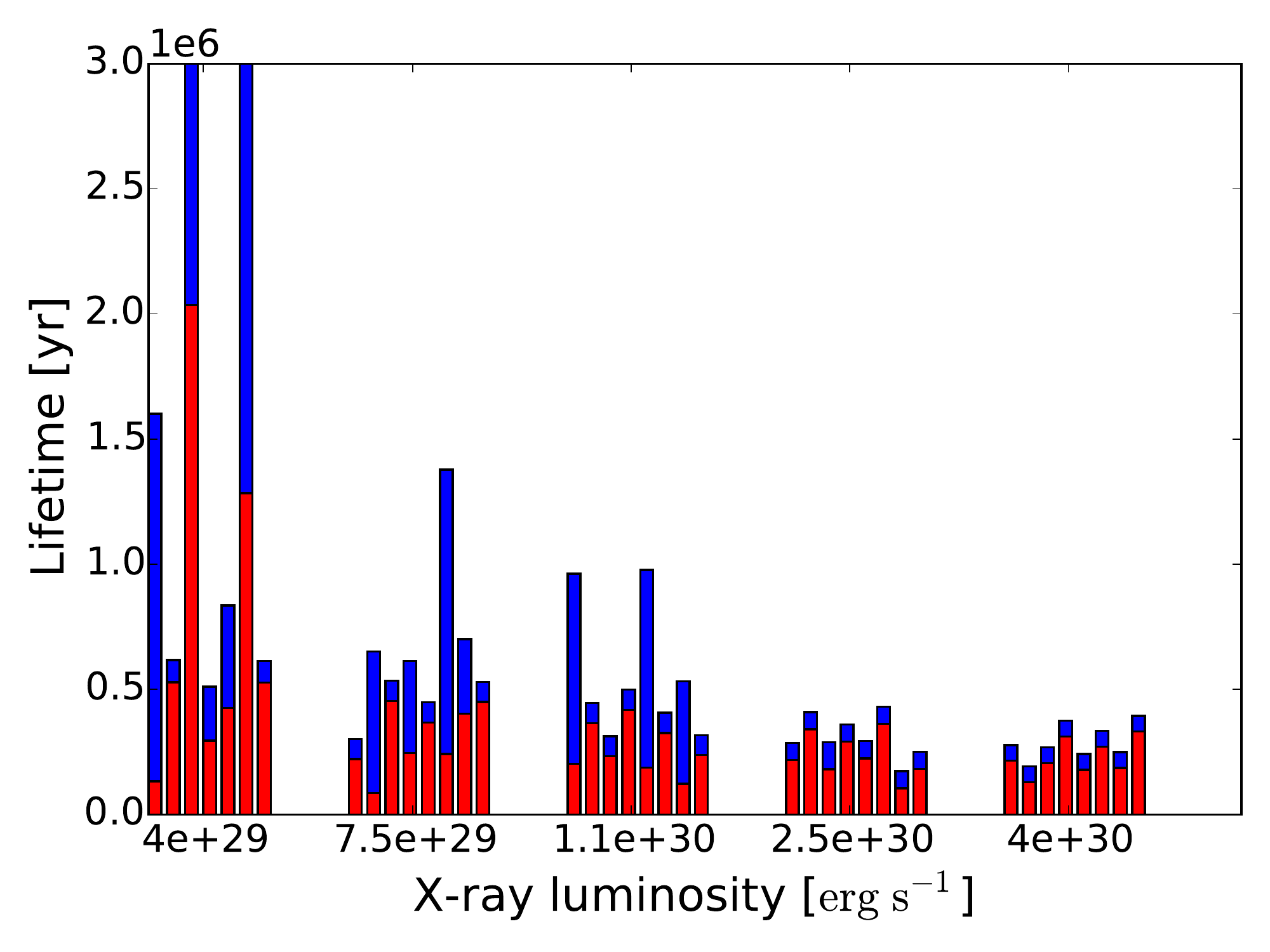}
\caption{Total lifetime of each simulation, decomposed in accreting (red) and non accreting phase (blue).}
\label{fig:lifetimes_bar}
\end{figure}

In figure \ref{fig:lifetimes_bar} we show a bar chart that decompose the life of each disc in the accreting and non accreting phase. The discs are grouped together based on their X-ray luminosity. The figure clearly shows the importance of X-ray luminosity on the total lifetime of the discs. The lifetime of the disc can range from a few $10^5$ years to several Myrs, and there is a significant trend for higher X-ray luminosities to yield shorter lifetimes. The average values are respectively roughly 1.5 Myrs, $6.5 \times 10^5 yr$, $5.5 \times 10^5 yr$, $3 \times 10^5 yr$ and $2 \times 10^5 yr$, from lowest to highest X-ray luminosity. This result is expected, since the photo-evaporation rates are higher for higher X-ray luminosities.

When looking at the relative length of the accreting and non accreting phases in figure \ref{fig:lifetimes_bar}, it can be seen that some discs almost never go through the non accreting phase. As soon as their outer disc is directly exposed to the X-ray radiation from the star, it is prone to the thermal sweeping instability. This depends very sensitively from the X-ray luminosity; we note that indeed all the discs we run at high X-ray luminosity go through this behaviour, as well as some of the ones at the median luminosity. In contrast, all the discs at low X-ray luminosity experienced a significant non accreting phase before being dispersed. This has important consequences on the expected population of transition discs found by observations. We can take, inside each X-ray luminosity bin, the fraction of discs that would be expected to be accreting. This is done by computing
\begin{equation}
f_\mathrm{acc} = \frac{\sum t_\mathrm{acc}}{\sum t_\mathrm{total}},
\end{equation}
that is, the mean duration of accretion divided by the mean lifetime. This calculation assumes that, save for the X-ray luminosity, the other two parameters have a uniform distribution. Although this is probably not true, their true distribution is not known; the purpose of the calculation is to get a handle on the result, and not to get an accurate statistical prediction. The fraction of accretors is found to be roughly $55$ percent for the first three X-ray luminosity values, and nearly $80$ percent for the last two cases. It is interesting that in the high X-ray luminosity case the fraction of accretors is quite high. Since only a handful of transition discs with large holes have been observed up to now (roughly 20 in the compilations of \citealt{Owen2012} and \citealt{Manara2014}), such a high fraction would explain why only few non accreting discs (for example, 2 in the compilation of \citealt{Manara2014}) have been observed. However, already in the median X-ray luminosity case the fraction goes down to one half. Clearly a better statistical treatment is needed to extract a predicted value, which we will discuss in the next section.

\begin{figure}
\includegraphics[width=\columnwidth]{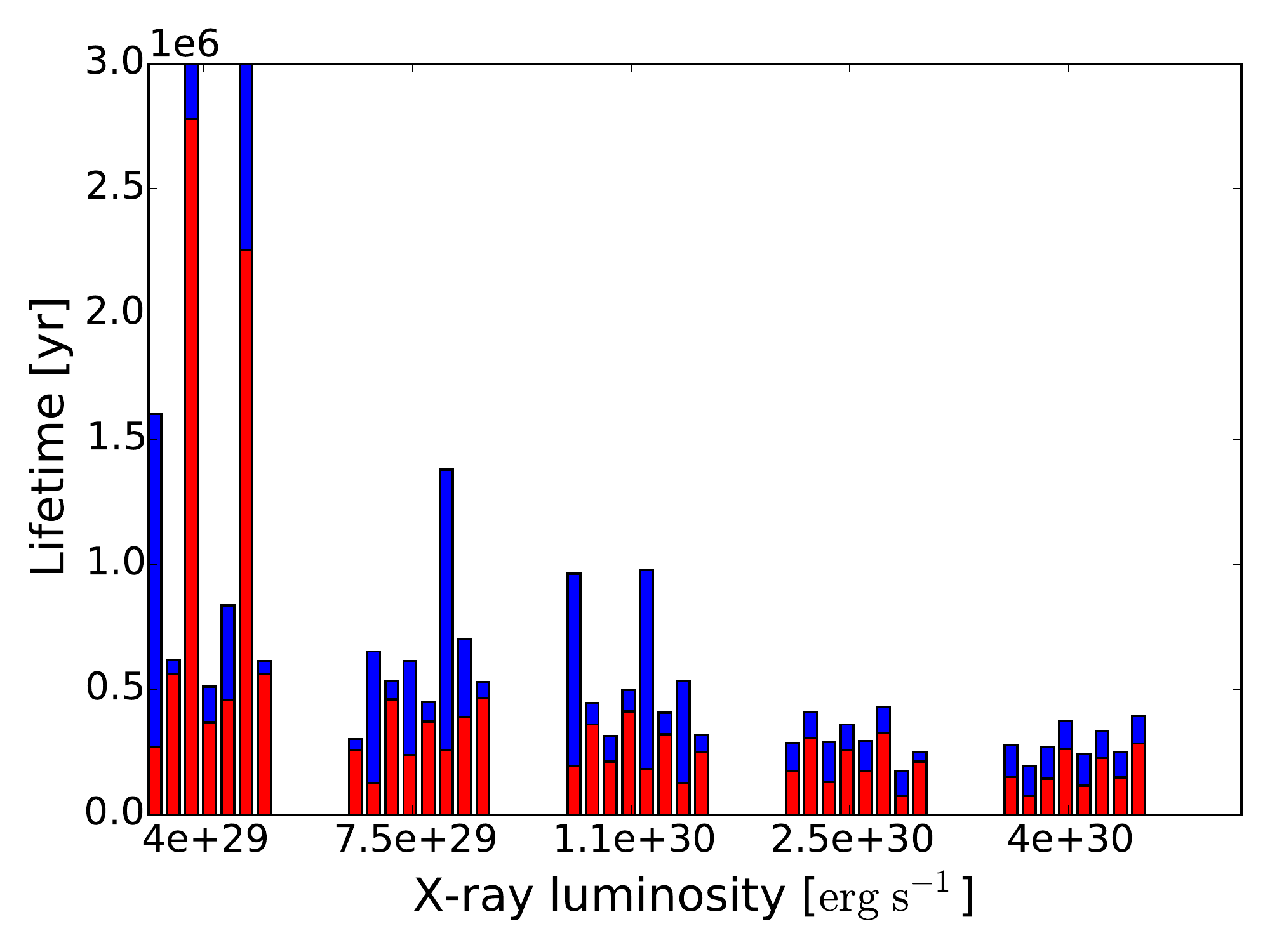}
\caption{Total lifetime of each simulation, decomposed in phase with (red) and without (blue) cold gas in the cavity}
\label{fig:lifetimes_bar_gas}
\end{figure}

Figure \ref{fig:lifetimes_bar_gas} shows again the lifetime of each simulations, but this time decomposed in the two phases with and without cold gas in the cavity. There is pratically no difference with the accretion for the median X-ray luminosity case; the fraction of discs expected to have cold gas in this case is 56 percent, which should be compared with the 57 percent of discs expected to be accreting. As explained in section \ref{sec:scheme}, the cold gas in the cavity and the presence of accretion have almost the same lifetime; some difference is due only to phase C, where the disc is still accreting but disc 2 has been already dispersed. However, phase C is so short that is practically unobservable. More interesting are the two cases with high X-ray luminosity; here the fraction of discs with cold gas in the cavity is respectively 64 and 58 percent, and thus the difference with the fraction of time spent accreting is more significant. This is due to the fact that all the other phases are faster with a high X-ray luminosity, so that the relative importance of phase C increases and the difference is larger. Although still unlikely given the number of observed transition discs, this predicts that one could in principle observe a disc that is still accreting, but no longer has cold gas in the cavity. Future observations with ALMA may be able to tell if such discs exist. Finally, also the lowest X-ray luminosity case is interesting. Here, the fraction of discs with cold gas rises to almost $70$ percent, which is puzzling to explain in the framework we presented up to now. The reason is that in this case there is a long phase where the disc is accreting (that is, it is still in phase A or B), but the accretion rate is under the current observational limits (we used $10^{-11} \ M_\odot \ \mathrm{yr}^{-1}$ as threshold), whereas ALMA will have the capability of detecting cold gas in the cavity down to very low masses. We thus classify such a disc as a non accretor, hence the lower time spent as an accretor compared to the time spent with gas in the cavity. As in the previous case, future observations will be able to tell if such discs exist.

\subsection{Observational prediction: the fraction of accreting discs}

We can use the results of our runs to estimate what would be the fraction of accreting transition discs found in observations. This involves creating a population synthesis model. We stress again that, for the purposes of this exercise, the term ``transition disc'' has a narrower meaning than what is normally meant. Here, it designates a disc containing a giant planet, and the purpose of this exercise is investigating whether we can explain the lack of non accreting transition discs with large holes. We generate a population of $N_\mathrm{disc}$ discs. We assign a random X-ray luminosity to these discs, distributed according to the X-ray luminosity function of Guedel (2007). We assume that a fraction $f_\mathrm{TD}=0.1$ of these discs become at a certain point of their life transition discs via the formation of a giant planet. We draw randomly the initial location of this planet between $20$ and $40 \ \mathrm{AU}$, and we also assign a random formation time $t_\mathrm{cr}$, uniformly distributed between $t_\mathrm{min}=10^5$ and $3 \times 10^6 \ \mathrm{yr}$. Note that we assume that there is no correlation between the formation time-scale and the X-ray luminosity, i.e. that the planet formation process is independent from the disc dispersal process. Since the assumed timescale of planet formation is of the same order as the disc lifetime (see introduction), some of our discs will be dispersed before they have a chance to form a planet. This highlights that, although the quantity that we are seeking to compute involves \textit{only transitional discs}, what the \textit{total} population of discs does is still relevant, as it selects which discs will form planets. In particular, this means that discs with higher X-ray luminosity are less likely to become transition discs via the formation of a planet. Being short-lived, they are more likely to have been already dispersed when they go in transition.

We  further use the results of our simulations to compute the timescales over which a newly formed transition disc will accrete and after which it will be dispersed by thermal sweeping. At any given time, we are thus able to compute how many discs have not been dispersed yet, how many are in transition, and how many of them are accreting. The key quantity that we want to compute from our models is the fraction of transition discs that we would expect to see accreting. The observations gather together discs that come from different star formation regions, rather than looking at a specific moment in time of a single region. In addition, it is commonly assumed that stars in the same region have an age spread between them (although the exact amount of this spread is debated). For this reason, we need to time average the fraction that we compute, under the assumption that the observations are sampling uniformly enough the timespan of disc lifetimes, and that there is a constant star formation rate in the neighbourhood of the Earth. Finally, we weight the fraction with the total number of discs that have not been dispersed yet, as observations are more likely to detect discs where there is more of them. In other words, we compute the time average of the number of transition discs that are still accreting, rather than the time average of the fraction. Mathematically, we compute the quantity
\begin{equation}
\frac{\displaystyle \int_0^{t_\mathrm{max}} S_\mathrm{disc} (t) f_{TD} (t) f_{TD,acc} (t)}{\displaystyle \int_0^{t_\mathrm{max}} S_\mathrm{disc} (t) f_{TD} (t) }
\label{eq:frac}
\end{equation}
where $S_\mathrm{disc} (t)$ is the survival function of the discs at time $t$ (that is, the fraction that is still alive), $f_{TD} (t)$ is the fraction of transition discs across the whole disc population, and $f_{TD,acc} (t)$ is the fraction of transition discs that are accreting. 

\begin{figure}
\includegraphics[width=\columnwidth]{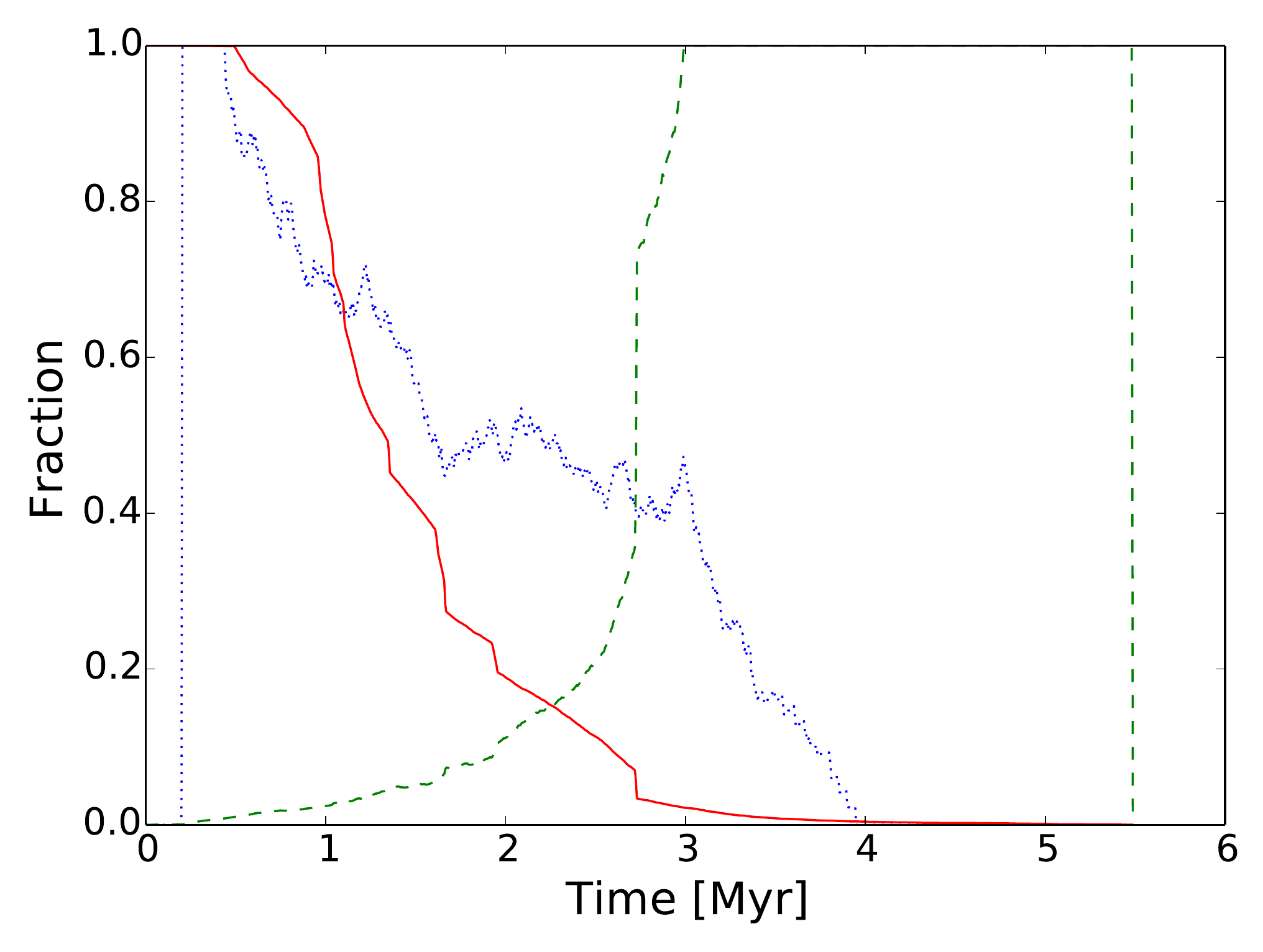}
\caption{The three quantities that appear in equation \ref{eq:frac}, namely $S_\mathrm{disc} (t)$ (red solid line), $f_{TD} (t)$ (green dashed line) and $f_{TD,acc} (t)$ (blue dotted line). }
\label{fig:MC_number_t}
\end{figure}

Figure \ref{fig:MC_number_t} shows the three quantities that appear in equation \ref{eq:frac} as a function of time, namely $S_\mathrm{disc} (t)$, $f_{TD} (t)$ and $f_{TD,acc} (t)$. The number of discs (red solid line) decreases roughly exponentially, as expected for a disc dispersal model. Conversely, the number of transition discs (green dashed line) increases with time: as planets form, some of the discs are converted into transition discs. Some of the transition discs formed can be very long lived, as phase D can be very long. The sharp cut-off around 6 Myr is a numerical consequence of the fact that we run the 2d hydro simulations only for 3 Myr at most. We note that this makes our estimate of the fraction of transition discs accreting an upper limit, and the predicted fraction should be even lower. Finally, the plot shows also the fraction of accreting transition discs (blue dotted line). This is not necessarily a monotonic function: as time passes, the number of already existing transition discs that are still accreting can only decrease, but new transition discs are formed, and all of them will initially be accreting. Overall though, the net effect is that the fraction is decreasing, that is, transition discs stop accreting at a faster rate than they are created.

\begin{figure}
\includegraphics[width=\columnwidth]{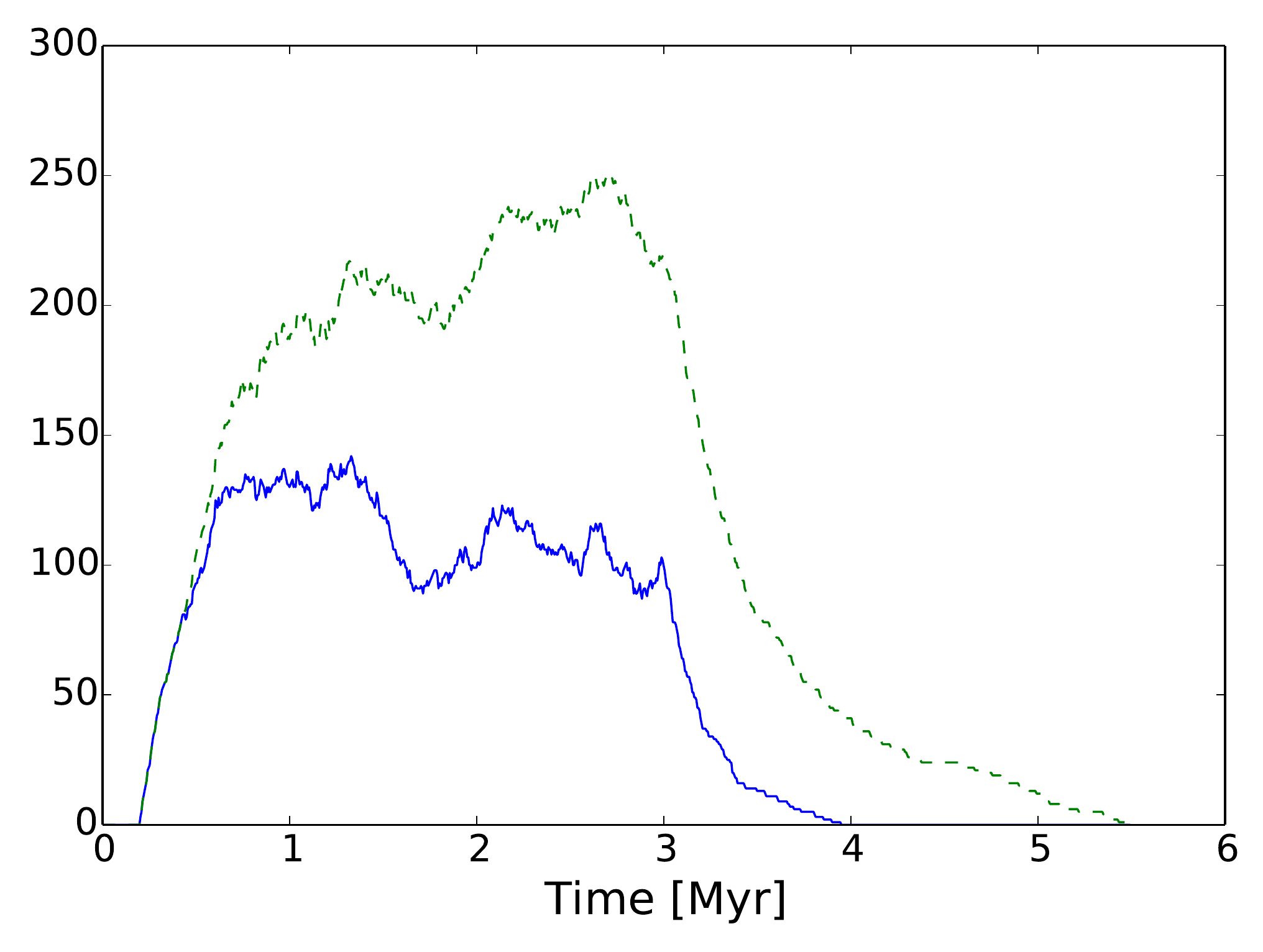}
\caption{The numerator (blue solid line) and the denominator (green dashed line) of Equation \ref{eq:frac} as a function of time, that is, respectively the fraction of accreting transition discs and the total number of transition discs (note that the normalisation is arbitrary).}
\label{fig:fractions}
\end{figure}

Figure \ref{fig:fractions} shows the numerator and the denominator of Equation \ref{eq:frac} as a function of time (note that the absolute normalisations of this functions are arbitrary). We note that the transition disc number is initially zero, as no disc has formed a planet yet. After $t_\mathrm{min}$ the discs start to be in transition, and for a while all of them are accreting. This can be seen by looking at the number of transition discs that are accreting, which initially overlaps with the other curve. The number of transition discs keep increasing until $3 \times 10^6 \ \mathrm{yr}$, when planets are not formed any more. Note also that after $\sim  5 \times 10^5 \ \mathrm{yr}$ some transition discs stop accreting, and the number of accreting transition discs is lower than the overall number of transition discs. After $\sim  4 \times 10^6 \ \mathrm{yr}$, no more transition discs are accreting, and after $\sim  6 \times 10^6 \ \mathrm{yr}$ all the transition discs have been dispersed. These absolute time-scales are obviously dependent on the assumed timescale of planet formation.

The final result of this section is that the time averaged value of the fraction of accretors is $\sim$ 0.48. Therefore, in our model we do not expect the majority of transition discs to be seen accreting. 

\section{Discussion and conclusions}

We have run 2d hydrodynamical simulations of photo-evaporating proto-planetary discs with a giant planet embedded. We have followed the evolution in time of such a system, which we called ``transition disc'', until the final dispersal of the outer disc by thermal sweeping. Our results show that the effectiveness of thermal sweeping in destroying the outer disc depends sensitively on the X-ray luminosity of the central star. When the X-ray luminosity is at the upper end of the X-ray luminosity function, the outer disc can be destroyed very rapidly. Otherwise, the outer disc is very long lived (even several Myrs).

To quantify better the implications of these results for observations, we have constructed a population synthesis model based on our simulations. We predict that only $\sim$ 48 $\%$ of transition discs should be observed accreting at rates greater than $10^{-11} \ M_\odot \ \mathrm{yr}$.

This is at odds with observations, where the great majority of transition discs with large holes are observed accreting above the mentioned rate. 
We discuss in the following points some of the assumptions used in this paper and possible solutions to reconcile the observations with the predictions of our model:
\begin{enumerate}
\item Discs with large holes, but no accretion are indeed present, but they have not been observed yet. This is similar to the idea of ``relic discs'' in the discussion of \citet{Owen2011}. If such discs showed no signatures of accretion and no infra-red excess above the stellar photosphere, they would be unlikely to be observed, and their host stars would be classified as diskless. However, even with holes of several tens of AU such as those we consider here the disc would have still significant emission at the Spitzer wavelengths of 24 and 70 $\mu m$. We checked that this is the case by inspecting the models in the \citet{robitaille2006} grid. Therefore we exclude this hypothesis. In addition, we note that recent ALMA observations \citep{Hardy2015} have placed strong upper limits on the masses of some relic discs, showing that they are severely depleted in gas mass. On the other hand, \citet{Espaillat2015} just added an object to the census of transition discs with large holes and no (or very low) accretion rate; the hole has been confirmed through sub-mm imaging. While this shows that some of these discs have gone unnoticed, it is unlikely that a population of lurking transition discs with large holes and very low accretion rates, as numerous as the current population of transition discs with large holes, is present.
\item It may be that "transition discs" created by carving of the disc by giant planets do not represent the last phase of proto-planetary disc evolution. Indeed, \citep{Andrews2011} find that the transition discs of their sample are more massive than classical T Tauri discs, possibly pointing out that they are younger, leading \citet{Owen2012} to argue that they maybe rare and long lived and not associated with disc dispersal. Note that our calculations just \textit{assume} that a disc with a planet is a transition disc, but this might not be the case. It may be that the inner hole phase caused by the planet is only temporary, and the disc may later then become a normal proto-planetary disc again before being eventually dispersed. This scenario fits nicely with the recently proposed idea by \citet{OwenRadPress} that transition discs are created by the accretion onto the planet will only appear as ``transition'' discs through a dip in the SED when the accretion rates are high $\sim10^{-8}$~M$_\ast$~yr$^{-1}$. At lower accretion rates \citet{OwenRadPress} argued that the planet was unable to trap small dust particles outside the planet's orbit and refilled the inner cavity. In this case one would expect to see a primordial SED, but still a large cavity in the mm as the mm-size particles are still trapped by the pressure maxima \citep{Pinilla2012}. However, although such a disc will show a conventional full SED once the mass accretion rate has decreased enough, in the framework presented in this work eventually photo-evaporation will remove the \textit{gas} inner disc (and therefore the dust with it), and the disc will be in what we called phase D. It would then be again a non-accreting transition disc with a large hole. For the detectability of such a source, see the previous point. So, while this process can explain why we do not see transition discs with huge cavities accreting at low rates, it does not explain why we do not see transition discs with huge cavities that are not accreting.
\item The planet induced photo-evaporation (PIPE) effect that we described in \citet{Rosotti2013} and in this paper for the case of X-ray photo-evaporating discs (see \citealt{alexander2009} that first described it in the context of EUV photo-evaporation) is largely due to the fact that the planet is accreting at high rates. This is necessary for example in the scenario of \citet{OwenRadPress} to push the small grains out of the gap and create a TD. More fundamentally, in the core accretion framework giant planets \textit{must} accrete gas at some stage in order to become giants. It might be speculated however that the accretion phase switches down after a while. Note that in standard core accretion this is not the case \citep{Pollack96}, as it is the initial phase that is planet limited, while after the collapse accretion is thought to be disc limited. On the other hand, if accretion went on at disc limited rates for a time comparable to the disc lifetime in a massive enough disc, all giant planets would end up having masses of several $M_j$ \citep{Szulagyi2014}. Instead, studies of exoplanets have shown that such planets are extremely rare (e.g. \citealt{OccurenceHARPS}). In hydrodynamical simulations like the ones that we employed here that lack the resolution to accurately model the flow inside the Hill sphere, this accretion rate is controlled by the parameter that sets the efficiency of accretion $f_\mathrm{acc}$ mentioned in section 2. If this parameter depends on time (or on the planet mass), so that it is reduced during the lifetime of the disc, the planet does not disturb so significantly the mass flow from the outer to the inner disc and PIPE is less effective in dispersing the inner disc. For example, in the extreme case where there is no accretion onto the planet, the presence of a planet does not make any difference to the evolution of a disc \textit{on viscous timescales} (e.g. \citealt{Zhu2011,duermann2015}), and therefore huge cavities are never opened in the gas. The dispersal of the inner disc would then happen when the disc is less massive, and thermal sweeping will take less time to disperse it. Note that this effect can coexist with the one described in the previous point, namely with a transition disc going back to a primordial phase, possibly offering a solution to the problem presented in this paper. From the observational side, it would be extremely useful to have measurements of accretion onto the putative planets present in transition discs that could inform the theory. Such searches are under way \citep{Close2014,Zhou2014} and it is likely that more data will arrive in the next future.
\item Photo-evaporation rates lower than the ones that we employed here would make the dispersal of the inner disc happen at later times. At this point, the mass of the disc would be lower, allowing thermal sweeping to set in on a quicker timescale. This could happen if, for example, the dominant heating mechanism is EUV photo-evaporation rather than X-rays. In addition, we note that the presence of huge cavities in the dust and of a gap created by a planet could affect the photo-evaporation rate. In this paper, we assumed rates computed for a full disc up to the moment of the inner disc dispersal. In particular, given that the cavities are depleted of dust (even in the small grains), the temperature of the disc will be severely affected. This will change the density vertical structure of the disc; the density at the launching point of the photo-evaporative wind, which sets the mass-loss rate, might be affected. Although X-ray photo-evaporation has been shown to be quite resistant to changes in the underlying disc structure \citep{Owen12Theory}, it remains to be seen whether this applies also to such extreme differences. In particular, even if this effect is unlikely to change the total mass-loss rate, it will probably change the flow topology \citep{Owen12Theory}, which might have implications for the inner disc. More work is clearly needed in this area.
\item Another unknown physical parameter that regulates the dispersal process is the value of $\alpha$ which sets the magnitude of the viscosity. Note that within a photo-evaporation model, this value is usually constrained by the requirement to reproduce the observed disc lifetimes. This is the reason why we did not let it vary in our simulations. Nevertheless, we can ask ourselves what would be the effect on the relative lifetimes of the accreting and non accreting phase. We can assume that the disc inside the orbit of the planet evolves according to a self-similar solution \citep{LyndenBellPringle,Hartmann98} where $\nu \propto r^\gamma$. For the purpose of this paper we assumed $\gamma=1$, but the value is not constrained and suitable values are in the range 0.5-1.5 \citep[e.g.,][]{WilliamsCieza2011}. If $\dot{M}_w$ is the photo-evaporation mass-loss rate from the inner disc, this disc is dispersed when the mass accretion rate onto the star becomes equal to $\dot{M}_w$. The duration of the accreting phase is thus:
\begin{equation}
\begin{split}
T_\mathrm{clear} &=t_\nu \left( \frac{\dot{M}_w}{ \dot{M}_0 } \right)^{-(2-\gamma)/(5/2-\gamma)} \\ &= t_\nu^{1/(5-2\gamma)} \left( \frac{\dot{M}_w}{ M_0 } \right)^{-(2-\gamma)/(5/2-\gamma)}
\end{split}
\end{equation}
where $\dot{M}_0$ is the initial mass accretion rate, $M_0$ the initial disc mass, and we have assumed $T_\mathrm{clear} \gg t_\nu$. This is a standard prediction of photo-evaporation models (e.g. \citealt{Clarke2001}). Thus $T_\mathrm{clear}$ scales with $\alpha^{-1/(5-2\gamma)}$, that is, for physically motivated values of $\gamma < 2.5$, more viscous discs are dispersed earlier, because the mass accretion rate evolves faster. The duration of the non-accreting phase is set by the time it takes to photo-evaporation to remove enough mass from the disc for thermal sweeping to kick in. In a self-similar solution, at any given time the mass of the disc $M_d \sim \dot{M} t$. Therefore the mass of the disc when the inner disc is cleared is roughly $M_d (T_\mathrm{clear}) \sim \dot{M}_w T_\mathrm{clear}$. If the mass needed for thermal sweeping $M_\mathrm{TS}$ is much smaller than this, the length of the non accreting phase is $T_\mathrm{NA} \sim M_d(T_\mathrm{clear})/\dot{M}_w$ (in reality $T_\mathrm{NA} \propto \Delta M=M_d(T_\mathrm{clear})-M_\mathrm{TS}$). Substituting the expression derived for $T_\mathrm{clear}$ we conclude that also $T_\mathrm{NA} \propto \alpha^{-1/(5-2\gamma)}$. To zero-th order thus the relative lengths of the accreting and non accreting phase do not depend on the value of the viscosity, so that it is difficult that a significant dependence on the value of viscosity is present. It is possible though that a dependence is present at higher orders or due to some of the assumptions that we made not being correct, which makes it warranting further study. In particular, while in the accreting phase we have considered only the viscous evolution as responsible for mass removal, actually also photo-evaporation will contribute. During the accreting phase photo-evaporation removes a mass of $M_\mathrm{lost} \sim \dot{M}_w T_\mathrm{clear}$, which lowers the mass at the start of the non accreting phase with respect to our previous estimate. This effect in particular is stronger for longer viscous time-scales, or for higher mass-loss rates (despite a shorter accreting phase). This partially explains why the discs with a higher X-ray luminosity have a very short non accreting phase, and points to the fact that the problem we described can be partially mitigated by invoking a lower value of the viscosity. In addition, if $M_d(T_\mathrm{clear}) \sim M_\mathrm{TS}$, the non accreting phase can be significantly shorter than we predicted. Again, this favours higher X-ray luminosities that have higher $M_\mathrm{TS}$, and is another reason why discs with high X-ray luminosities do not suffer from the problem of long non-accreting phases.

\item  Up to now we assumed that the planet does not migrate. Migration could be attractive from a theoretical point of view, as it could decrease the sizes of the cavities in transition discs, while at the same time the disc is decreasing in mass and mass accretion rate onto the star. If migration is fast enough, it can have reduced significantly the cavity size by the time PIPE dissipates the inner disc (that is, phase B starts) and shuts down accretion. This scenario of migrating giant planets in transition discs has been considered by \citet{clarke2013}. However their conclusions were that planet migration is actually too fast compared to the decrease in the disc mass, so that, unless unreasonably high planet masses are considered, the planets would rapidly reduce their separation from the star. This would create transition discs with small cavities, but which are still massive and therefore bright in the sub-mm, which are not observed \citep{Owen2012}. We note that recent results \citep{duffell2014,duermann2015} on Type II (giant) planet migration imply that planet migration could be even faster than what assumed in \citet{clarke2013}, exacerbating even more this problem. To understand if migration is important for the models presented in this paper, we can do a check \textit{a posteriori} by comparing the timescale for migration with the lifetime of the disc. For example the timescale for planet migration quoted by \citet{clarke2013}, assuming a mass accretion rate of $10^{-8} \ M_\odot \mathrm{yr}$, is a few Myr. This is indeed long compared to the lifetime of our discs (see figure \ref{fig:lifetimes_bar}). Note however that their disc masses are a factor of a few higher, which would make the disc lifetime correspondingly longer. To remain within our models, the viscous time-scale at 20 AU is $\sim 8 \times 10^5 \ \mathrm{yr}$, which is marginally longer than the lifetime of most of our discs. However, note also that the planet accretes a significant amount of material during the lifetime of the simulation, so that its final mass can become as big as several $M_j$. In this case, type II migration is in the so-called ``planet-dominated'' regime \citep{SyerClarke1995}, and will be therefore slower than the pure viscous time-scale of a factor $\sim M_p/\pi \Sigma r^2$. For example, at 20 AU, for the \textit{initial} planet mass of $0.7 M_J$, having type II migration on a viscous time-scale requires a surface density of $\sim 5-6 \ \mathrm{g} \ \mathrm{cm}^{-2}$. This is significantly more than the surface density at which thermal sweeping sets in (see equation \ref{eqn:ths}), meaning that there will be a long phase where migration is planet-dominated. Thus, photo-evaporation can efficiently disperse the disc before the planet has migrated significantly in most cases. However, for the longest lived discs in our set migration might be an important effect.

\end{enumerate}

Perhaps the most important lesson to be learnt is that we should focus not only on the mechanism to create the massive transition discs seen in observations, but also on their long term evolution. Very few models have attempted to tackle this problem so far. The experiment conducted in this paper consisted in coupling the presence of a planet to models of photo-evaporation, which we know is happening in discs (as mentioned in the introduction). Bearing in mind the caveats and open points discussed above, the combination of the two is not capable of explaining the lack of non accreting transition discs with large holes. We stress that this is currently an open problem, that no model has addressed so far, and is a {\it serious} issue in the planetary hypothesis for transition discs. We are left to speculate that we are missing one of the ingredients operating in transition discs. Whatever the mechanism responsible for the dispersal of transition discs is, it must act on a very quick time-scale to explain the apparent lack of non-accreting sources. Moreover, taking our calculations as a \textit{reductio ad absurdum} would lead us to conclude that one of the hypothesis that we have assumed leads to a contradiction. As there is evidence that discs are subject to photo-evaporation, this poses strong constraints to scenarios in which a massive planetary companion is responsible for the creation of a transition disc, or, at least, in which it is the \textit{only} culprit. Similar conclusions were reached by \citet{clarke2013}. We also remark that, from the observational side, it would be useful to have more constraints on the ``descendants'' of these transition discs, as for example ALMA is doing for evolved discs in general \citep{Hardy2015}, and on their ``precursors''. It is likely that in this process we will gain also more insight on the processes leading to the formation of transition discs and shed more light on their nature.

Future work could investigate the effect of the inclusion of other energy sources, such as the FUV that is important at large separations from the star, in the gas heating. This might enhance the efficiency of thermal sweeping and thus produce a higher fraction of accretors.





\label{sec_concl}


\section*{Acknowledgements}
We thank the referee, Richard Alexander, for the constructive criticism that improved this paper. Giovanni Rosotti acknowledges the support of the International Max Planck Research School (IMPRS). We acknowledge the support by the DFG Cluster of Excellence "Origin and Structure of the Universe". JEO acknowledges support by NASA through Hubble Fellowship grant HST-HF2-51346.001-A awarded by the Space Telescope Science Institute, which is operated by the Association of Universities for Research in Astronomy, Inc., for NASA, under contract NAS 5-26555. The simulations have been carried out on the computing facilities of the Computational Center for Particle and Astrophysics (C2PAP). This work has been supported by the DISCSIM project, grant agreement 341137 funded by the European Research Council under ERC-2013-ADG. We thank Marco Tazzari for discussions about the statistics of transition discs. We thank Veronica Roccatagliata, Cathie Clarke and the rest of the DISCSIM group, and the ESO star formation coffee for interesting discussions.

\bibliography{BiblioRosotti}{}
\bibliographystyle{mn2e}

\bsp

\label{lastpage}

\end{document}